\documentclass[lettersize,journal]{IEEEtran}
\usepackage{amsmath,amsfonts}
\usepackage{algorithmic}
\usepackage{algorithm}
\usepackage{array}
\usepackage[caption=false,font=normalsize,labelfont=sf,textfont=sf]{subfig}
\usepackage{textcomp}
\usepackage{stfloats}
\usepackage{url}
\usepackage{verbatim}
\usepackage{graphicx}
\usepackage{xcolor}
\usepackage{cite}
\begin{document}

\title{Design of Cryogenic Fully Differential Gain Boosting-OTA by the $g_{m}/I_{d}$ methodology used for a 14 bit Pipelined-SAR ADC}

\author{Mingjie Wen, 
	Chao Luo,
	BoLun Zeng and 
	Guoping Guo}
\maketitle

\begin{abstract}
Quantum computing (QC) requires cryogenic electronic circuits as control and readout sub-systems of quantum chips to meet the qubit scale-up challenges.At this temperature,MOSFETs transistors exhibition many changes such as higher threshold voltage,higher mobility,and steeper substhreshold slope.We present a cryogenic fully differential gain boosting-OTA used for a 14 bit Pipelined-SAR ADC operating at 4.2K as the readout circuit for semiconductor-based quantum computing system.Using $g_{m}/I_{d}$ methodology to get pre-computed lookup tables based on the cryogenic 110nm BSIM4 model.The proposed OTA achieves very high unity-gain frequency@1.23GHz and open-loop low frequency gain@101dB.The total power consumption is 2.66mW at 4.2K,and a setting accuracy better than 0.01\% with $f_{-3dB}$ of 37MHz in a closed-loop application.
\end{abstract}


\section{Introduction}\label{secI}

\IEEEPARstart{Q}{uantum} processors based on arrays of quantum bits (qubits) including semiconductor quantum dots or superconducting transmons operate at sub-Kelvin cryogenic temperatures.Bringing the quantum-classical interface closer to the qubit is the core solution for overcoming the qubits scale-up challenge. Classical electronic components will be great changed even invalid in such a condition and the use of MOSFETs operating at cryogenic temperatures (cryo-CMOS) has been stydied extensively\cite{charbon201715}.
Cryogenic MOSFETs' modeling plays a key role in the control and readout circuits for quantum computing.MOSFETs' cryogenic characters have been widely reported in recent years\cite{beckers2018characterization}\cite{zhang2021hot}\cite{beckers2018design}.
In a general way,higher threshold voltages, higher mobility and steeper subthreshold swing(SS) can be found in CMOS at cryogenic temperature.In this paper,a commercial 110-nm low threshold voltage MOSFETs (LVT-MOSFETs) technology is experimentally characterized at liquid helium temperature(LHT) and room temperature(RT).
Based on that,we have builded a BSIM4 compact model which can be used in commercial SPICE or HSPICE  simulator for verifying cryo-CMOS analog circuits.Using this accurate model,We have designed a folded cascade gain-boosting operational transconductance amplifier(OTA) by the $g_{m}/I_{d}$ method used for a cryogenic 14 bit Pipelined-SAR analog to digital convert(ADC).

$g_{m}/I_{d}$ methodology is a powerful philosophy for analog designers especially for the nanoscale MOSFET.The 110nm Cryo-CMOS is complicated and its IV-behavior can't match the classic square law model,which is based on ideal dirft model and applies only near the strong inversion(SI)\cite{jespers2017systematic}.For this reason,we use pre-computed lookup tables to maintain a model-agnostic approach.The goal of this work is to design a high performance and low power consumption OTA by the $g_{m}/I_{d}$ method based on a 110nm cryo-CMOS BSIM4 model.The fully differential gain boosting-OTA  works at 4.2K and is used as a residue amplification in a two-stage Pipelined-SAR ADC.

This paper is divided into five sections.The second section will introduce the characteristics of the 110nm LVT-CMOS at RT and LHT and the third section focuses on the $g_{m}/I_{d}$ method,includes the establishment of pre-computed lookup tables.The fourth section  focuses on the high performance and low power consumption OTA,and the fully differential gain boosting structure and anaylysis of the circuit.The fifth section will give the post simulation of the OTA,the performance characteristics at RT and LHT.

\section{Cryogenic CMOS}

\begin{figure}[htb]
	\centerline{\includegraphics[width=9cm]{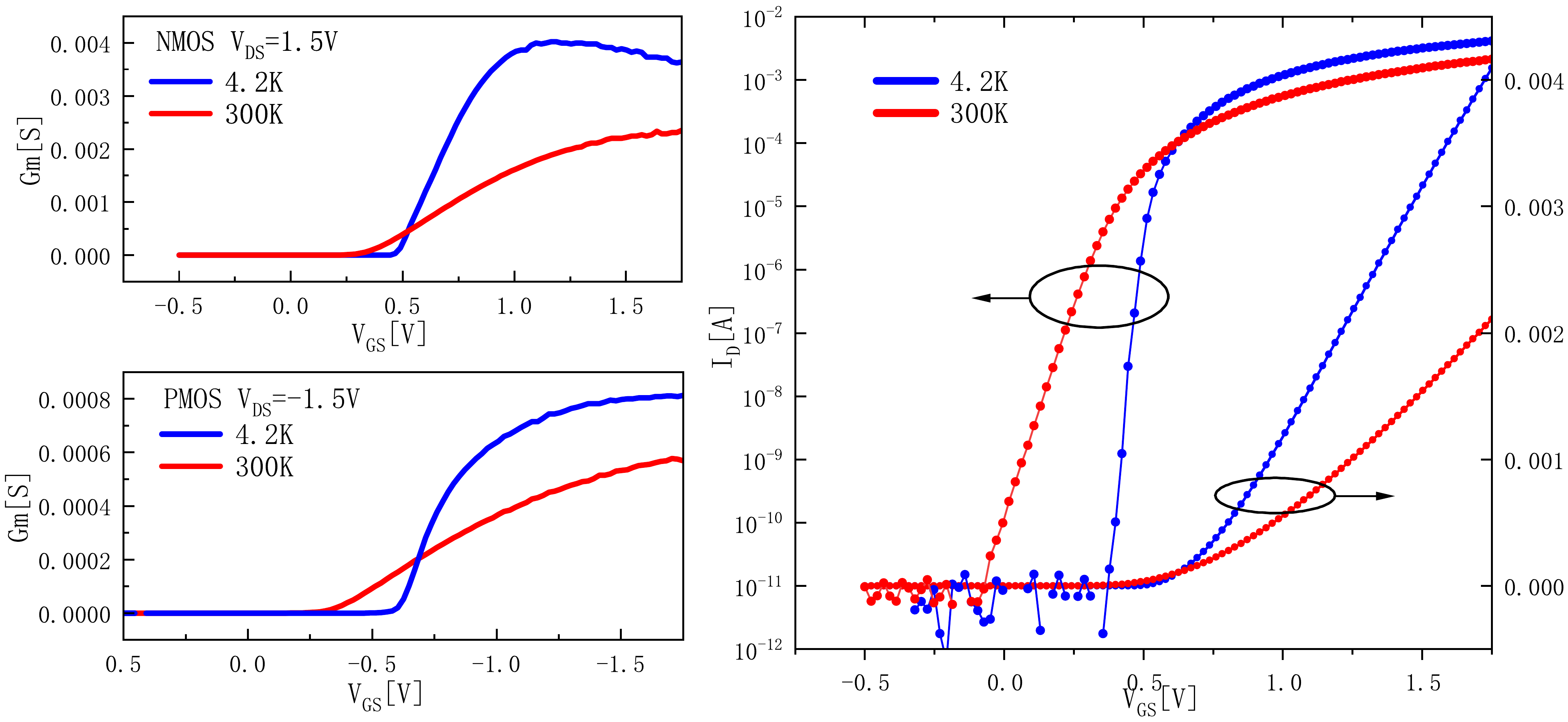}}
	\caption{gm of NMOS and PMOS }
	\label{fig:gm_id_vg}
\end{figure}

MOSFETs have specific phenomenons at cryogenic temperature,such as kink effect,impurity freeze-out effect and so on.As 110nm model shows,the low-field mobility increases almost ten times for the 10$\mu m /10\mu m$  size NMOSFETs at LHT.As illustrated in Fig.\ref{fig:gm_id_vg}(a,b),the MOSFETs' $g_{m}$ increases two or three times at LHT,which means great advantages for analog circuits design 
comparing to RT.But because of the substrate Fermi potential increasing at cryogenic temperature, a higher threshold voltages $V_{th}$ can be found in  Fig.\ref{fig:gm_id_vg}(c). Also,we find that the ameliorated ON/OFF ratio increases at LHT. a steeper subthreshold swing(SS).

SS is the most important parameter in low-power design and it's proportional to the slope factor n and the thermal voltage $U_T(kT/q)$. 
$$SS=n U_T ln10$$
Usually n lies between 1.2 and 1.5 for bulk technology at RT,and is somewhat smaller in silicon on insulator(SOI) transistors. At 4.2K temperature,SS predicts to be 71 times smaller than RT,but because of the change of the  density of the interface-traps,SS reduces to only 2-3 times smaller than RT. 

We have characterized 110nm LVT-MOSFETs at RT and LHT and have bulided the BSIM4 model based on these commericial LVT devices.These device under test(DUT) have lower Vth than conventional devices,so we can get a fast switching speed even though Vth increases at LHT.According to Fig.\ref{fig:gm_id_vg}(c),we extracted Vths of the DUTs: Vthn=0.33734V and Vthp=-0.35077V at RT,which change to 0.4356V and -0.6861V at LHT respectively using maximun transconductance method.In weak inversion(WI) region:
$$\frac{g_{m}}{I_{d}}=\frac{1}{n U_{T}}=\frac{ln10}{SS}$$
At LHT,we can find that the SS decreases two or three times than RT,so we got highter $g_{m}/I_{d}$ at LHT.That's important for low-power analog/RF circuit design.


\section{$g_{m}/I_{d}$ design method} 

%
In the WI region,the current is due to diffusion and mostly due to drift in SI region; $V_{OV}$ has played a central role in classical square-law design,it determines which region the MOSFETs in.But this simple model is obsolete with modern nanoMOSFET transistors.Modern EKV and BSIM transistor models are extremely complicated,they usually have thousands parameters and the basic first order IV-behaviour can't match the simulation results. 


Some reasearchs present that analog circuit designers can use the inversion coefficient IC for classifing operation regions of a MOSFET\cite{beckers2018design}.

\begin{equation}
	\centering
	\begin{cases}
		IC<=0.1&\text{weak inversion(WI)} \\
		0.1<IC<=10&\text{moderate inversion(MI)}\\
		IC>=10&\text{strong inversion(SI)}
	\end{cases}
\end{equation}
and
$$IC = \frac{I_{D}}{I_{spec}}|_{saturation} = \frac{I_{D}}{2nU_T^2\mu C_{ox}W/L} = \frac{L}{{2nU_T^2\mu C_{ox}}}J_{D}$$
and the $I_{spec}$ is the product of specific currents per square $I_{spec_{\square}}$ and $W/L$.But the IC parameter is still connected with the  technology viriables.It's not the best solution for cryogenic analog design to maintain a model-agnostic approach.

We use $g_{m}/I_{d}$ as a proxy,it determines the MOSFET's inversion level.It also controls some most important Figure-of-Merit that analog circuit designers care about.
$$\frac{g_{m}}{I_{d}}=\frac{1}{I_{D}}\frac{\partial I_D}{\partial V_G}=\frac{2}{V_{OV}}$$
For long-channel MOSFETs,the $g_{m}/I_{d}$ is invariant of the technology.Short channel effect(SCE) degrade $g_{m}/I_{d}$ because of the velocity saturation\cite{enz2015low}

\begin{figure}[htb]
	\begin{center}
		\centerline{\includegraphics[width=9cm]{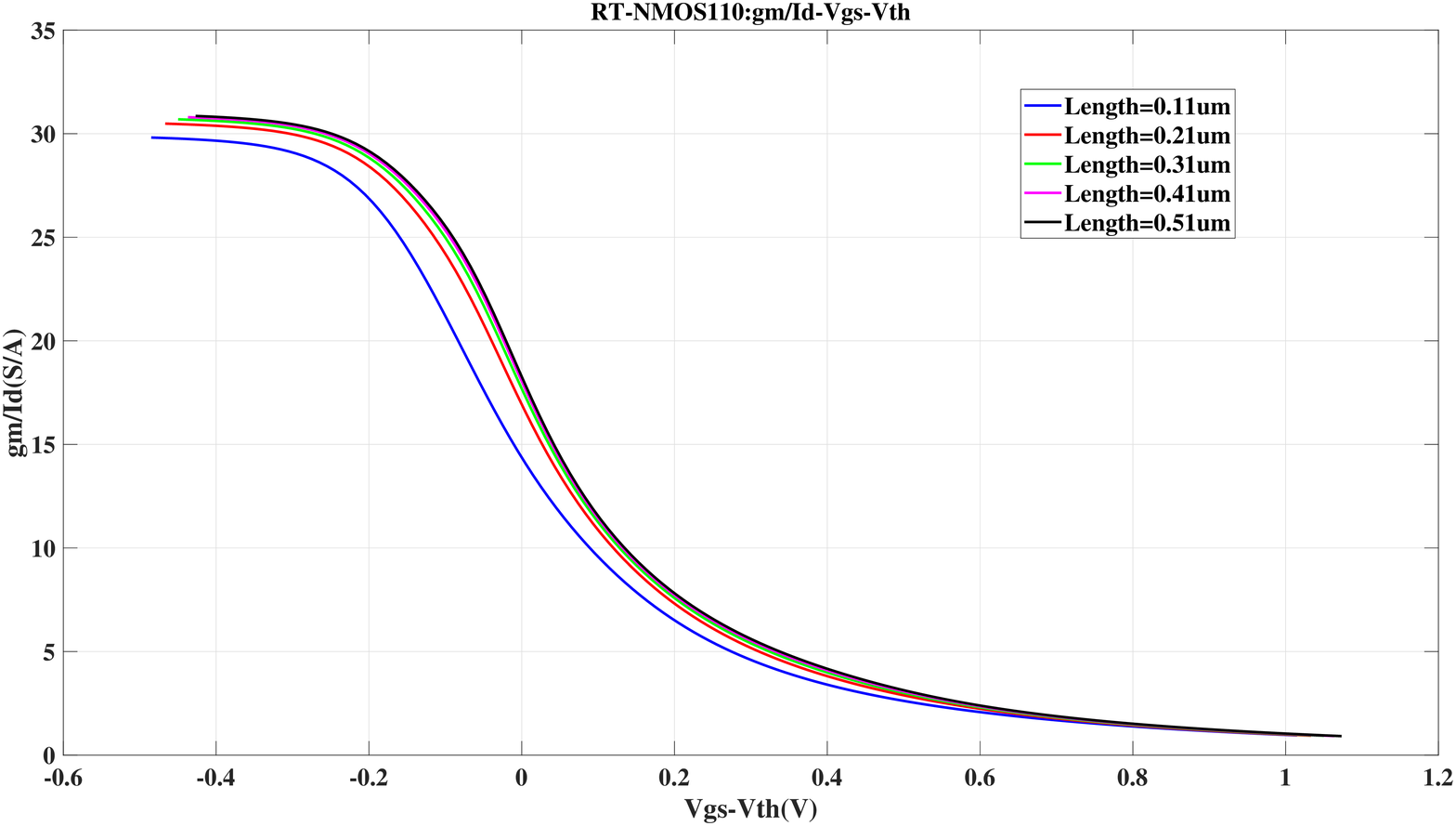}}
		\centerline{\includegraphics[width=9cm]{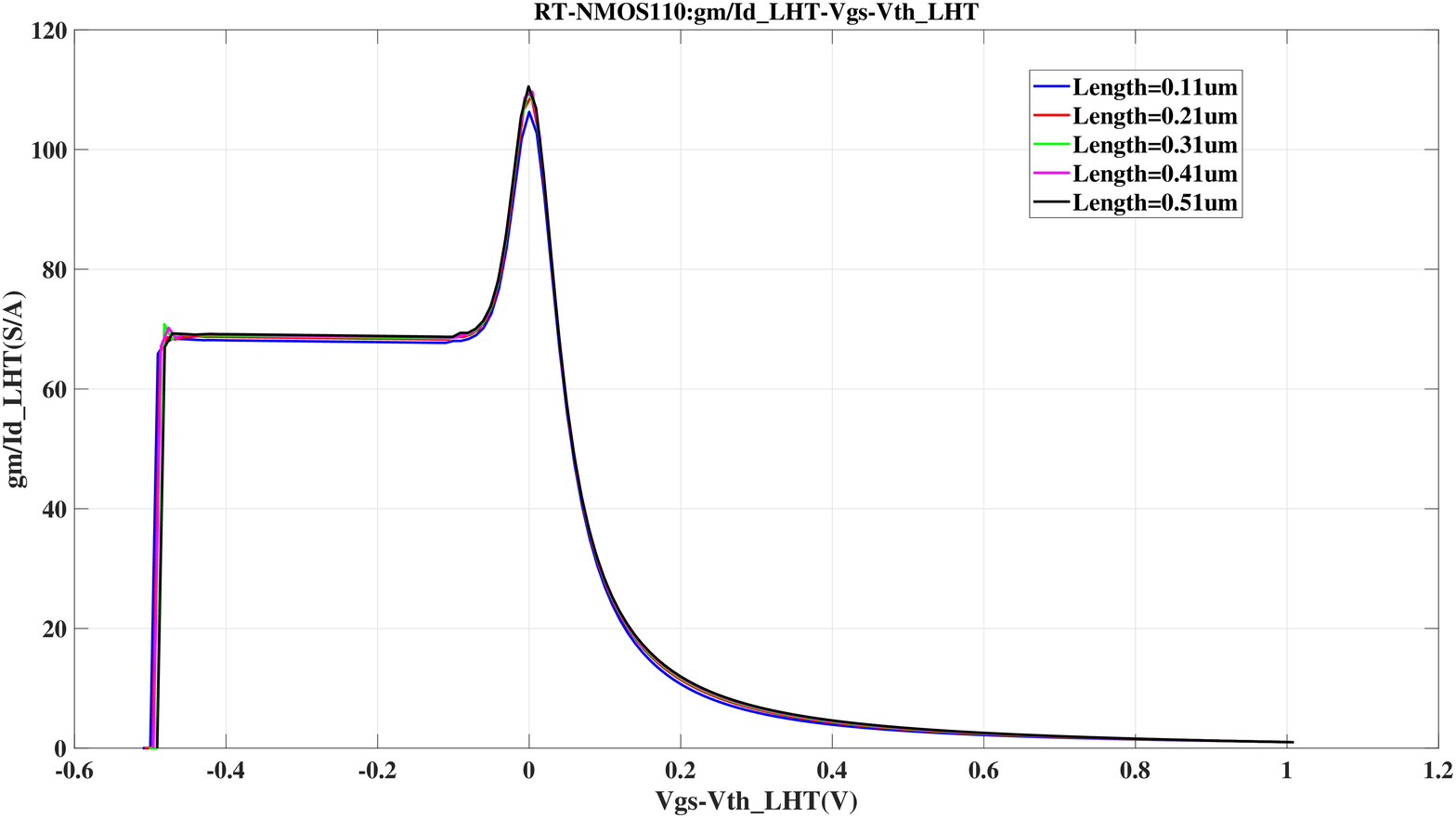}}
	\end{center}
	\caption{300k and 4k temperature gm/id and  Vgs-Vth}
	\label{fig:gmid_vdsat }
\end{figure}

As a beginning ,we tabulate some parameters ,such as transit frequency$f_{t}(g_{m}/2\pi C_{gg})$,intrinsic gain$g_{m}/g_{ds}$,capacitance scale$C_{gd}/g_{gg}$,$C_{dd}/g_{gg}$and current density$J_{D}(I_{d}/W)$ for different channele lengths and $g_{m}/I_{d}$.All these variables are width-independent quantities.
In the WI level,$g_{m}/I_{d} = 1/nU_{T}$
and the lowest power consumption can be write as followed:
$$I_{dmin}=\frac{g_{m}}{(g_{m}/I_{d})_{max}}=g_{m}\cdot nU_{T}$$
Fig.\ref{fig:gmid_vdsat }(a,b) shows that the MOSFETs get a higher $g_{m}/I_{d}$ under the same inversion level at 4.2K temperature. In other word, cryogenic analog  circuits  may consume less static power than room temperature operating at WI region or MI region.

\begin{figure}[htb]
	\begin{center}
		\centerline{\includegraphics[width=9cm]{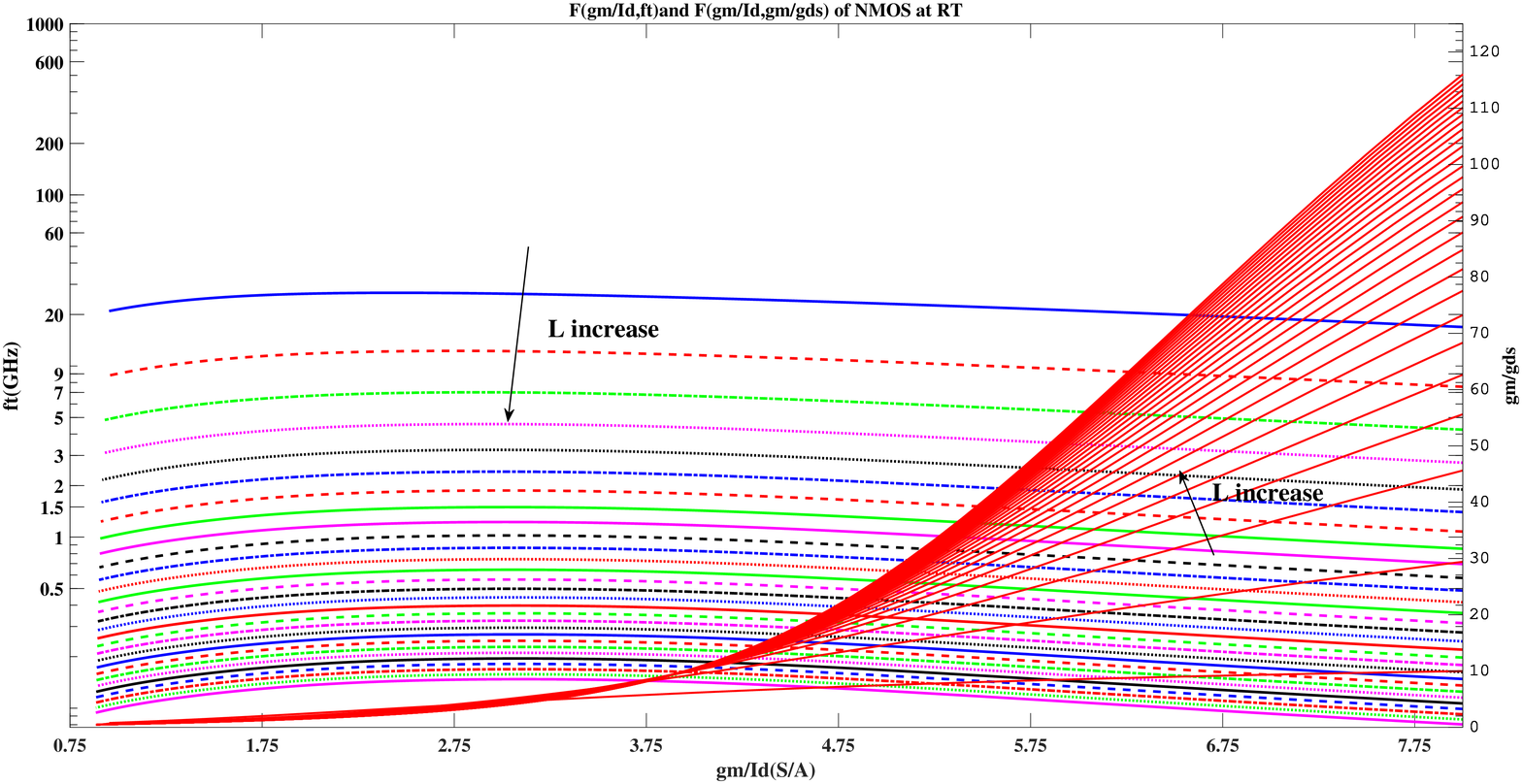}}
		\centerline{\includegraphics[width=9cm]{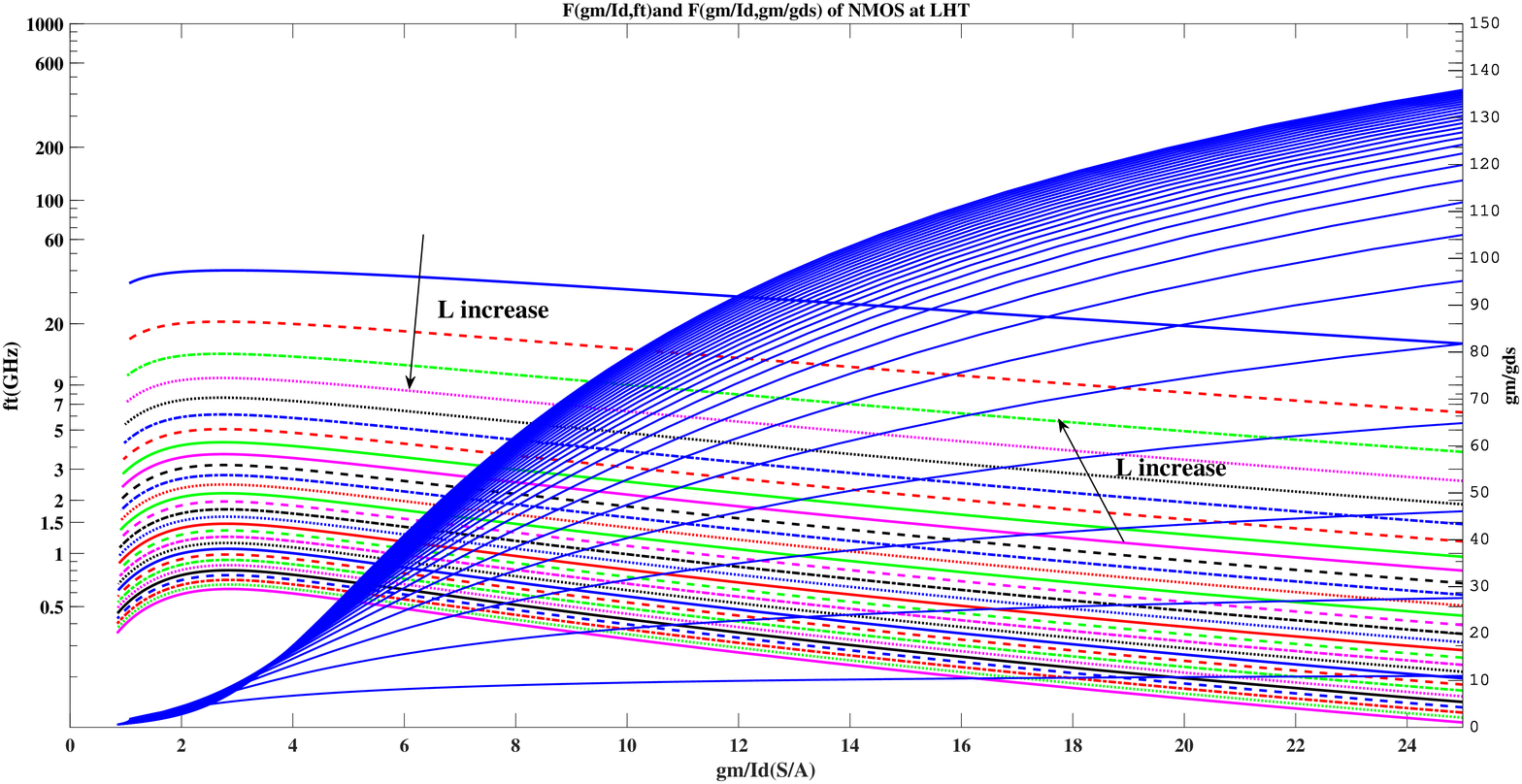}}
	\end{center}
	\caption{300k and 4k temperature gm/gds\_gm/id and ft\_gmid }
	\label{fig:gm/gds and ft vs gmid }
\end{figure}

Fig.\ref{fig:gm/gds and ft vs gmid }(a,b) shows two most important figures of merit of  n-channel device.It shows the relationship of $g_{m}/g_{ds}$ VS $g_{m}/I_{d}$ and $f_t$ VS $g_{m}/I_{d}$ at 300K and 4.2K.

According to the unity-gain frequency(GBW) requirement,we get the $g_{m1,2}=2\pi \cdot GBW\cdot C_{L}$,then pick a reasonable channel length L, for a short channel we get the high $f_{T}$,and for a long channel,we get high intrinsic gain $g_{m}/g_{ds}$ .Then pick the $g_{m}/I_{d}$ value,for a large value,we can save the power and the $V_{OV}$ will be small and get large signal swing. According to the requirement of SR and $C_{L}$,we get the current value,check the lookup tables of the $I_{d}/W$ and $g{m}/I_{d}$,we will get the width of the transistor.
In another way ,we can determine $I_{d}$ from $g_{m}$ and $g_{m}/I_{d}$,then determine W from $I_{d}/W$.

\section{Proposed OTA}
\begin{figure}[htb]
	\centerline{\includegraphics[width=9cm]{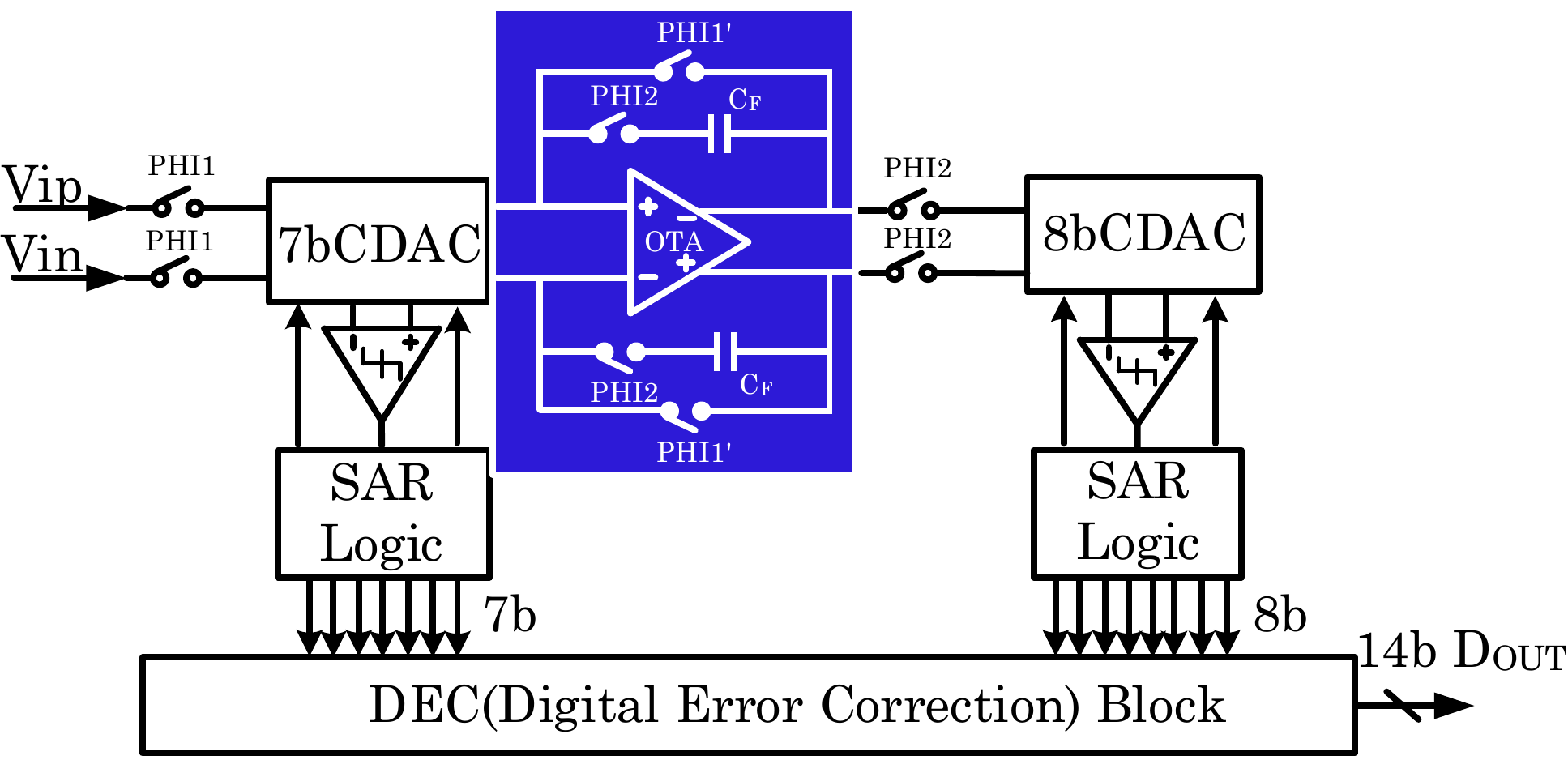}}
	\caption{the 14 bit Pipelined-SAR ADC structure and residue amplifier}
	\label{fig:14 bit Pipelined-SAR ADC structure}
\end{figure}

Based on the 110nm cryo-CMOS BSIM4 model,we have designed a high performance and low power consumption OTA by the $g_{m}/I_{d}$ method used for a 14bit Pipelined-SAR ADC.This ADC is used as a readout subsystem for the semiconductor QC platform,which includes a 7bit SAR ADC, a residue amplifier and a 8bit SdAR ADC.As illustrate in  Fig\ref{fig:14 bit Pipelined-SAR ADC structure}, the first 7b ADC quantify the input analog data as the first seven bits,and the residue is amplified by the closed-loop residue amplifier to the second 8bit SAR ADC.Then the residue is quantified as the second eight bits.A digital error correction(DEC) module converts the 15bit digital datas to the final 14bit digital codes as final output.

This OTA is based on a 110nm BSIM4 model and need to satisfy the following  specifications:
\begin{itemize}
	\item[$\bullet$] power supply  AVDD=1.5V
	\item[$\bullet$] operating temperature 4.2K
	\item[$\bullet$] open-loop gain $A_{V0}$$>$84.5dB
	\item[$\bullet$] unity gain frequency GBW $>$993MHz
	\item[$\bullet$] capacitive load C$_{L}$$>$2.3pF
\end{itemize}


The nondominant poles lie in so high frequency that telescopic and folded-cascode topologies have faster speed than two-stage structrue,and they can be just considered as core candidates.Folded-cascode accommodate a wider input common mode voltage range,take into concideration of the high GBW,we choose folded-cascode as the main OTA sturcture.In this project,we use NMOS as input transistor,though it has lower second nondominant pole than PMOS input,it has a higher input $g_{m}$ than PMOS under same current,which is the key solution for saving power consumption.
the nondomanating pole $P_{2}$ can be write as:$$gm_{5,6}/{2\pi (C_{DD1,2}+C_{DD3,4}+C_{GS5,6}+C_{par\_p\_booster})}$$

\begin{figure}[htb]
	\begin{center}
		\centerline{\includegraphics[width=9cm]{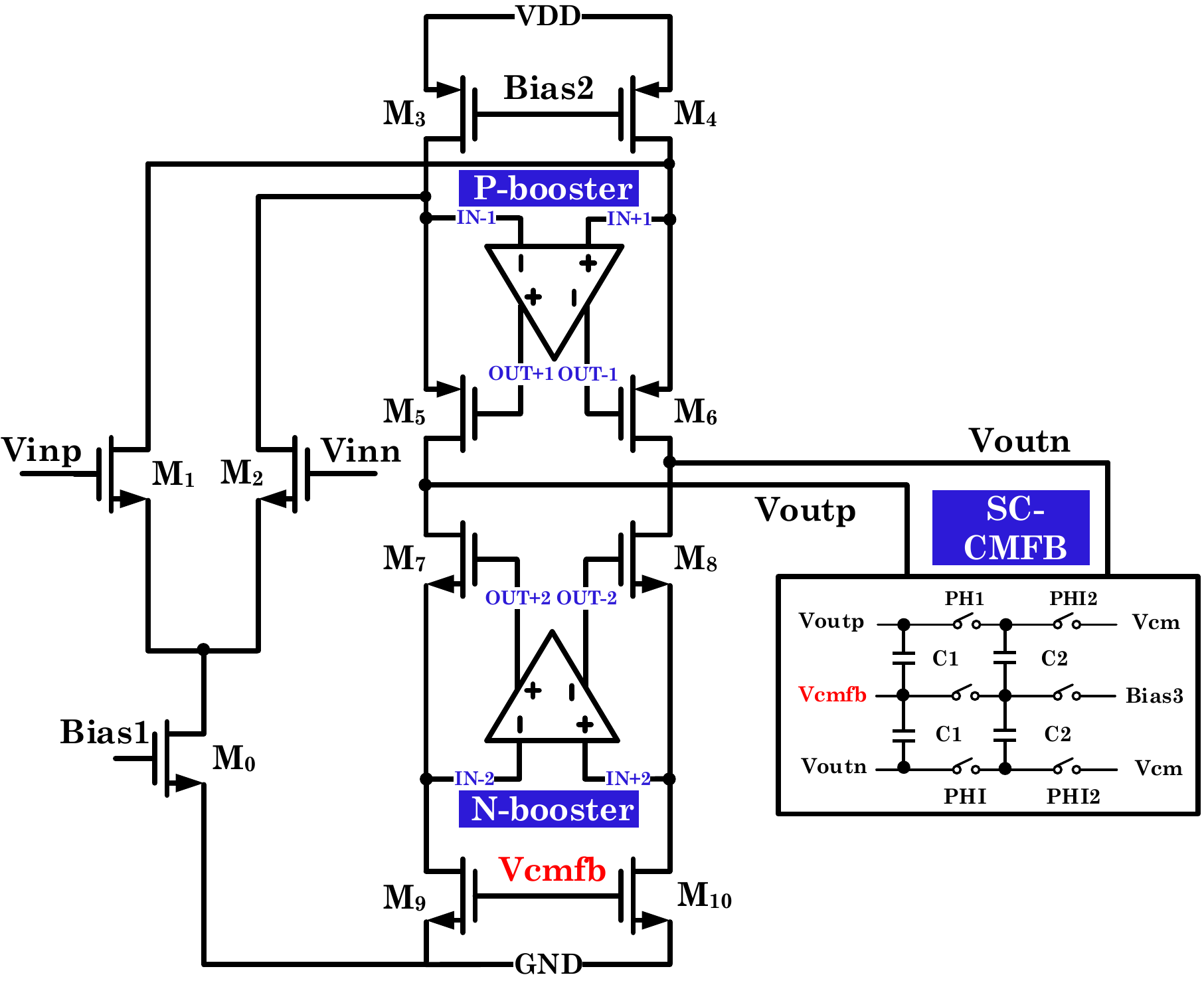}}
	\end{center}
	\caption{Struture of the gain boosting-OTA }
	\label{fig:gain_boosting structure}
\end{figure}

$C_{par\_p\_booster}$ means the P-booster amplifier's input parasite capacitance.
Fig.\ref{fig:gain_boosting structure} show the structure of the main fully differential folded-cascode OTA circuit.Using this gain-enhancement technique we can get a even bigger $Z_{out}$ and the total gain is 
$$g_{m1,2}\{((1+A_{bp})g_{m5,6}r_{o5,6}r_{o3,4})||((1+A_{bn})g_{m7,8}r_{o7,8}r_{o9,10})\}$$

$A_{bp}$ and $A_{bn}$ means the open-loop gain of p-booster and n-booster amplifiers.

\begin{figure}[htb]
	\begin{center}
		\centerline{\includegraphics[width=9cm]{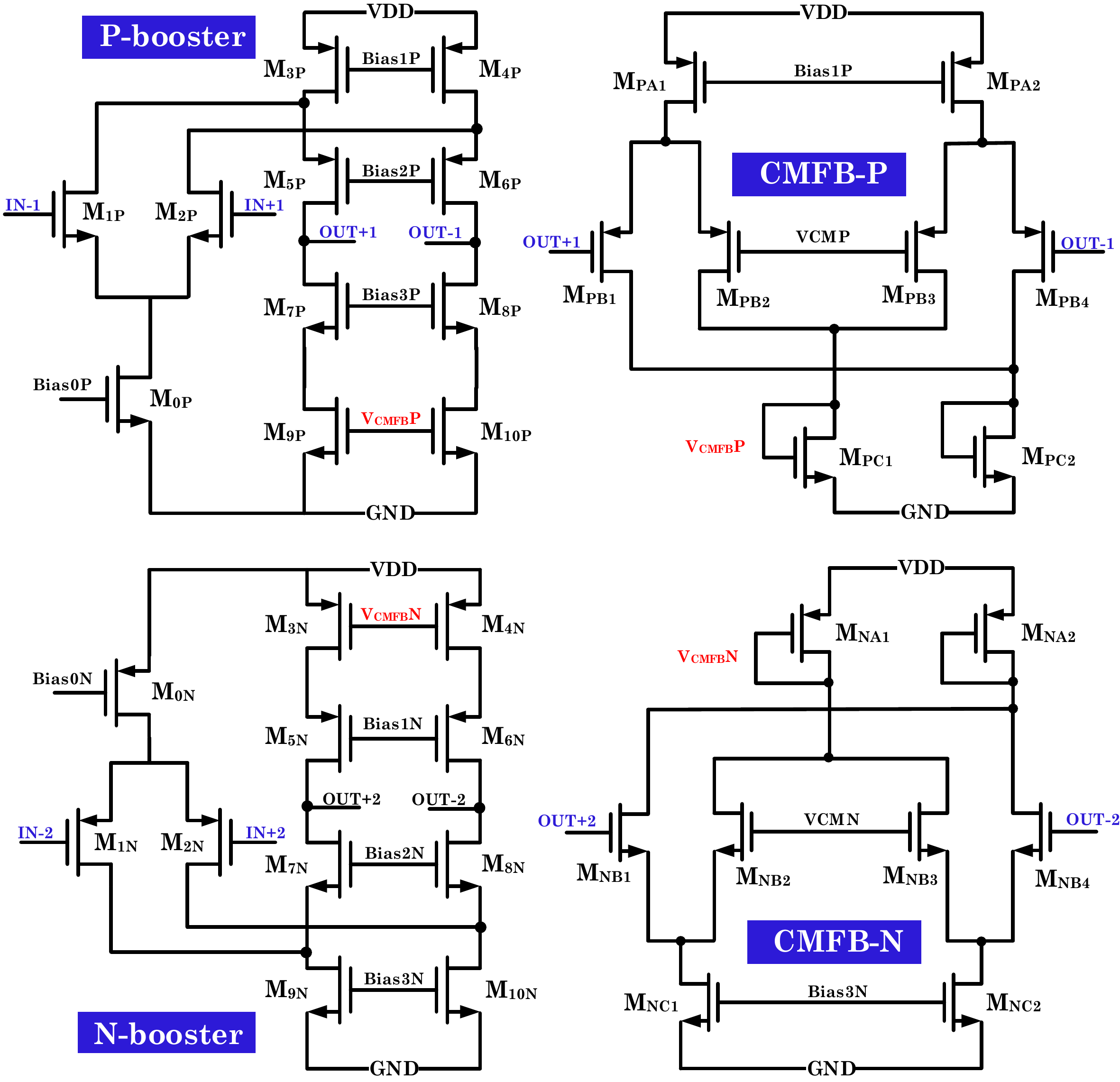}}
	\end{center}
	\caption{booster structure }
	\label{fig:npboosting}
\end{figure}

There are two non-overlapping clocks in the design of Pipelined-SAR ADC,so the CMFB modual of the main OTA can use the switch capacitance for convenience.
When PHI2=1,the quantity of electric charge is
$$Q_{1}=2C_2(V_{cm}-V_{bias3})+C_{1}(V_{outp}+V_{outn}-2V_{cmfb})$$
and when PHI1=1,the quantity of electric charge is
$$Q_{2}=(C_{1}+C_{2})(V_{outp}+V_{outn}-2V_{cmfb})$$
Let $Q_{1}=Q_{2}$,we realize the purpose of extracting the common mode voltage$ (V_{outp}+V_{outn})/{2}$ and comparing it with a known $V_{cm}$.
$$V_{cmfb}=\frac{V_{outp}+V_{outn}}{2}-V_{cm}+V_{bias3}$$

As for the structure of booster ampifiers,we can take the p-booster as an example.Set input common mode voltage is Vpi,and output common mode voltage is Vpo.In the main folded cascode OTA we can get $$Vpi=V_{DD}-V_{DS3,4}$$
$$Vpo=V_{pi}-V_{GS5,6}$$
and
$$Vpo=Vpi-|Vthp|-V_{OV5,6} $$
In view of this limiting condition,the basci 5-T and telescope structure  cannot be choosed as the solution for booster amplifiers.

Fig.\ref{fig:npboosting} demonstrates the structure of the p-booster and n-booster amplifier and the CMFB circuits.It's fully differential folded cascode  as same as the main OTA.Because of the input voltages of the booster amplifiers change little,so we choose the continue time-CMFB circuits.To avoid the doublets slowing the setting time casused by the p-booster and n-booster amplifiers,The unity-gain frequency of the booster amplifiers must be in the range of $f_{-3dB}$ of the closed-loop amplifier and the nondominant pole $P_{2}$ of the main OTA\cite{bult1990fast}.
$$\beta \omega_{m} < \omega_{booster} < \omega_{P_{2}} $$

$\beta$ is the feedback factor and $\beta \omega_{m}$ is the closed-loop  $f_{-3dB}$ bandwith. $\omega_{P_{2}}$ the first nondominant pole of the main OTA.


\section{Post simulation results}

\begin{figure}[htb]
	\begin{center}
		\includegraphics[width=8cm]{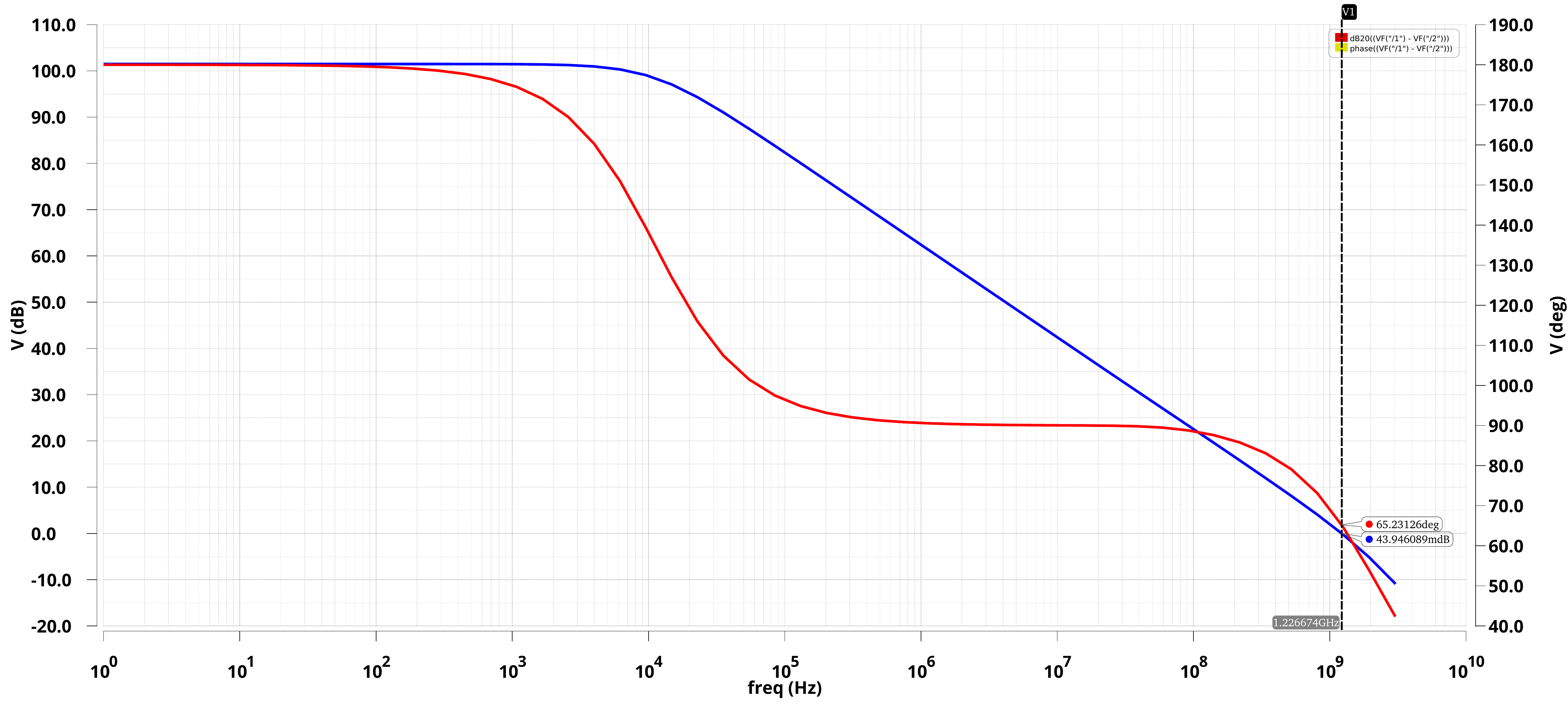}
	\end{center}
	\caption{Bode\-plot simulation of the Open loop gain and GBW of the proposed  OTA}
	\label{fig:final_bode_plot measurement}
\end{figure}

Bode-plot measurements show a open-loop gain of 101dB and a unity-gain frequency $\omega_{m}$ of 1.23GHz(2.3pF load) in Fig.\ref{fig:final_bode_plot measurement}.And the power consumption is as low as 2.77mW.Setting measurements with a feedback factor $\beta$ of 1/33 show a fast single-pole settling behavior behavior correspondign to a closed-loop $f_{-3dB}$ of 37MHz and a setting accuracy better than 0.01\%.


\section{Conclusion}

\bibliographystyle{IEEEtran} 
\bibliography{ref}

\vfill

\end{document}